# Threshold Activation Reaction and Absorption Dose Rates Inside and on the Surface of a Thick W-Na Target Irradiated with 0.8 GeV Protons

Yu.E. Titarenko, V.F. Batyaev, E.I. Karpikhin, V.M. Zhivun, A.B. Koldobsky,
R.D. Mulambetov, S.V. Mulambetova, V.E. Luckjashin, K.A. Lipatov, S.G. Mashnik [1], R.E. Prael [1]

Institute for Theoretical and Experimental Physics (ITEP), B.Cheremushkinskaya 25, 117259 Moscow, Russia
[1] Los Alamos National Laboratory, Los Alamos, NM 87545, USA

**Abstract** *Results are presented of measuring the threshold activation reaction rates in $^{12}$C, $^{19}$F, $^{27}$Al, $^{59}$Co, $^{63}$Cu, $^{65}$Cu, $^{64}$Zn, $^{93}$Nb, $^{115}$In, $^{169}$Tm, $^{181}$Ta, $^{197}$Au, and $^{209}$Bi experimental samples placed both along the axis inside and outside a 0.8 GeV proton-irradiated thick W-Na target. Absorbed dose rates outside the target are presented as well. The target was irradiated by the proton beam from the ITEP U10 accelerator. The proton fluence and proton beam shape were monitored by the $^{27}$Al(p,x)$^{7}$Be reaction. 158 reactions of up to ~0.5 GeV thresholds have been measured in 123 activation samples. The reaction rates were determined using the γ-spectrometry method. The absorbed dose rates were measured via 10 samples made of silver-enriched glass using technique of stimulated photo-luminescence. In total, more than 1000 values of activation reaction and dose rates were determined in the experiment. The measured reaction rates were compared with the LAHET code simulated rates and using several nuclear databases for the respective excitation functions, namely, ENDF/B6 for cross section of neutrons at energies below 20 MeV and MENDL2 together with MENDL2P for cross sections of protons and neutrons of 20 to 100 MeV energies. A general satisfactory agreement between simulated and experimental data has been found. Nevertheless, further studies should be conducted for the purposes of perfecting the simulation of the production of secondary protons and high-energy neutrons, especially in the backward direction with respect to the beam. The results obtained permit some conclusions concerning the reliability of the transport codes and data bases used to simulate the ADS with Na-cooled W targets. The high-energy threshold excitation functions to be used in the activation-based unfolding of neutron spectra inside the ADS can also be inferred from the results.*

## I. INTRODUCTION

The pending researches with the pilot Accelerator Driven Systems (ADS) require reliable nuclear data. One of the possible fields of the data application is the activation-based unfolding of the high energy (up to ~ 1 GeV) neutron spectra inside the ADS target and the near-target blanket zone. The high-energy "tail" in ADS spectra triggers the high-energy threshold (>10 MeV) reactions, which are not studied in detail in the conventional reactor researches. Therefore, the excitation functions that may be derived from high-energy calculations or retrieved from the available databases strongly require reliable verification via testing experiments with the high-energy proton-irradiated target micromodels.

In some applications, the space distributions of the secondary radiation-induced dose near the principal units of ADS facilities are essential. In particular, the distributions define such an important parameter of the ADS facilities as their service life.

The ITEP U-10 proton synchrotron was used to realize a run of experiments to study the threshold activation reaction rates and the absorbed doses inside and outside a thick W-Na target.

## II. EXPERIMENT

The target was assembled of alternating cylindrical W and Na discs (see Fig. 1). The displayed succession of the disks was selected to obtain the maximum attainable uniform neutron distribution along the target.

The W discs are the 150-mm diameter full-metal structures of 40-mm (7 pieces) and 20-mm (5 pieces) thicknesses. The discs were prepared of a WNiFe alloy (97.5% W, 1.75 Ni, 0.75% Fe) by the powder metallurgy techniques. To be irradiated inside the target, the experimental samples were fastened to the job-oriented rulers, which were inserted tight into the grooves slotted in the discs. Fig. 2 shows the samples arrangement on W disks.

The Na discs (13 pieces) are the 150-mm diameter, 40-mm thick thin-walled cylindrical containers made of 0.5-mm thick stainless sheet steel, filled with metallic Na, and soldered tight.

The target having been assembled, the discs were fixed in a job-oriented cradle. After that, the cradle was taken to the irradiation position to be aligned there in such a manner that the cylinder axis would coincide with the proton beam axis.

The experimental activation samples are 10.5-mm or 10.0-mm diameter circular discs cut of metal foils of certified composition. During the irradiation runs, the samples were placed on the W disc surfaces and in their seats milled in the rulers (see Fig. 2). In the experiments, the activation reaction rates were determined on the $^{209}$Bi, $^{197}$Au, $^{169}$Tm, $^{115}$In, $^{93}$Nb, $^{65}$Cu, $^{64}$Zn, $^{63}$Cu, $^{59}$Co, $^{27}$Al, $^{19}$F, and $^{12}$C nuclides. Besides, the $^{22}$Na production rate in the Na discs was determined.

The target was irradiated with the ITEP synchrotron-extracted 0.8 GeV proton beam at ~0.25s$^{-1}$ pulse repetition rate and ~0.5s single pulse duration. The irradiation lasted for 10 h. Within the irradiation run, the beam intensity was monitored using an integral ionization chamber.[1]

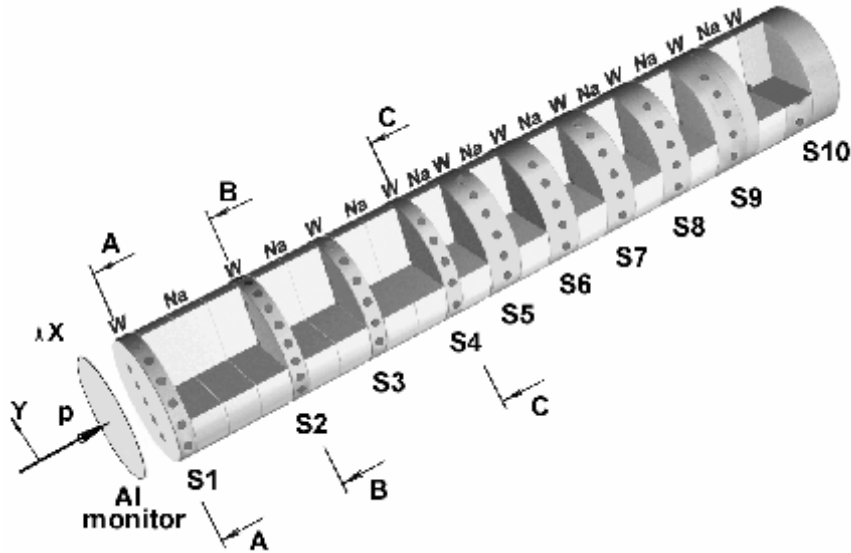

Fig 1. The W-Na target assembly view.

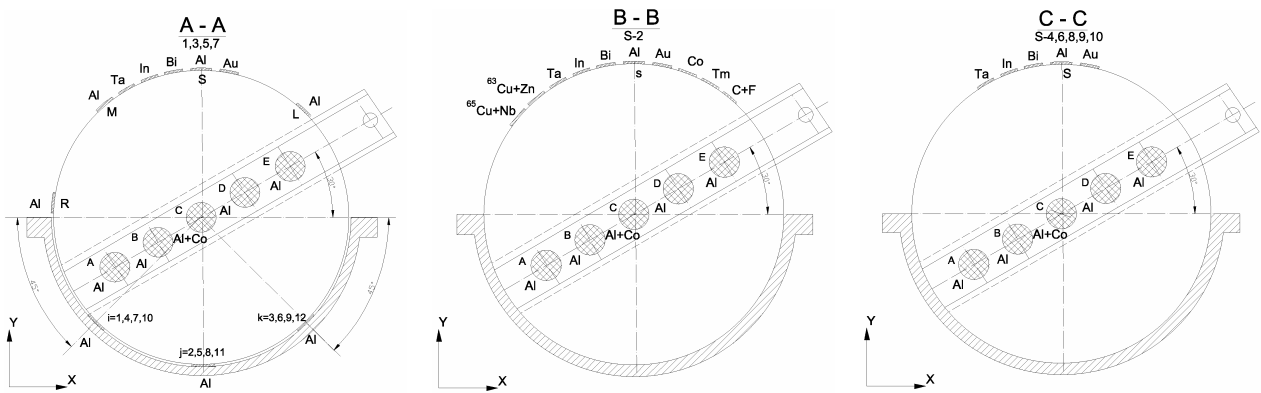

Fig. 2. Arrangement of experimental samples on W discs. The A-A sectional view shows discs Nos. 1, 3, 5, and 7; the BB view shows disc No. 2; the C-C view shows discs Nos. 6, 8, 9, and 10.

The number of protons that hit the W-Na target together with the proton beam shape was determined using a 150-mm diameter Al monitor placed across the beam at 50 mm in front of the first disc. Having been irradiated, the monitor was cut into fragments, and the numbers of the $^7$Be, $^{22}$Na, and $^{24}$Na product nuclei were determined in each fragment. The monitoring was realized mainly for $^7$Be, and for $^{22}$Na and $^{24}$Na as an ancillary procedure. The $^{27}$Al(p,x)$^7$Be reaction cross section for 0.8 GeV protons was taken to be 6.4+-0.4 mb.[2] The final proton number on the target calculated taking this cross section value proved to be $(6.5+-0.4)*10^{14}$.

The irradiated activation samples were γ-spectrometered. Use was made of the Ge and Ge-Li detector-based spectrometers of 1.8-keV and 2.5-keV resolutions, respectively, in the 1332.5 keV $^{60}$Co γ-line. The spectrometers have been calibrated for measurements at a few fixed source-detector distances within the 40-375 mm range. In the measurements, the sample-detector distances were selected basing on the admissible count rate of the spectrometer.

The above presented version of the techniques for processing the results of the γ-spectrometric experiment, allowing for the pulsed mode of the accelerator, is described in.[2,3] The final formulas to determine the experimental reaction rates in the activation samples are presented in.[4] For the experimental reaction rates to be compared correctly with the calculation results, they were normalized to the proton beam power $W$, which was determined as $W=N_p \cdot E_p/T$ (where $N_p$ is the proton number on the target; $E_p$ is the proton energy; $T$ is the irradiation time) and proved to be (2.32+-0.15) W.

The reaction rate errors were calculated using the standard error transfer formulas. The uncertainties in the γ–yields and monitor reaction cross section make the major contribution to the errors.

In total, 979 values of the reaction rates have been determined in the experiment (see Table 1).

Table 1. List of the reaction rates measured

| Nuclide | The products for the reaction rates measured | Number of values |
|---|---|---|
| $^{209}$Bi | $^{207}$Po, $^{206}$Po, $^{206}$Bi(i,c), $^{205}$Bi, $^{204}$Bi, $^{203}$Bi, $^{203}$Pb(i,c), $^{201}$Pb, $^{200}$Pb, $^{202}$Tl, $^{200}$Tl(i,c), $^{191}$Pt, $^{96}$Tc, $^{96}$Nb, $^{82}$Br | 165 |
| $^{197}$Au | $^{198}$Au, $^{196}$Au, $^{194}$Au, $^{191}$Pt | 33 |
| $^{181}$Ta | $^{182}$Ta, $^{178m}$Ta, $^{176}$Ta, $^{175}$Ta, $^{173}$Ta, $^{180m}$Hf, $^{175}$Hf, $^{173}$Hf, $^{170}$Hf, $^{172}$Lu, $^{171}$Lu, $^{169}$Lu, $^{167}$Tm, $^{165}$Tm, $^{161}$Er | 112 |
| $^{169}$Tm | $^{166}$Yb, $^{168}$Tm, $^{167}$Tm, $^{166}$Tm(i,c), $^{165}$Tm, $^{163}$Tm, $^{161}$Er, $^{160}$Er, $^{160m}$Ho(i,c), $^{157}$Dy, $^{155}$Dy, $^{153}$Dy, $^{152}$Dy, $^{153}$Tb, $^{152}$Tb, $^{151}$Tb, $^{150}$Tb | 19 |
| $^{115}$In | $^{113}$Sn, $^{116}$In, $^{115m}$In(i,c), $^{114m}$In, $^{113m}$In(i,c), $^{111}$In, $^{110}$In, $^{109}$In, $^{115}$Cd, $^{111}$Ag, $^{110m}$Ag, $^{106}$Ag, $^{105}$Ag, $^{101}$Pd, $^{100}$Pd, $^{101}$Rh, $^{100}$Rh(i,c), $^{99}$Rh, $^{97}$Ru, $^{96}$Tc, $^{95}$Tc, $^{90}$Mo, $^{99}$Nb(i,c), $^{89}$Zr, $^{87}$Y | 207 |
| $^{93}$Nb | $^{90}$Nb, $^{89}$Zr, $^{86}$Zr, $^{90}$Y, $^{87m}$Y, $^{87}$Y, $^{86}$Y(i,c) | 8 |
| $^{65}$Cu | $^{64}$Cu, $^{61}$Cu, $^{65}$Ni, $^{58}$Co, $^{57}$Co, $^{56}$Co, $^{55}$Co, $^{56}$Mn, $^{52}$Mn, $^{44m}$Sc, $^{44}$Sc(i,m+g) | 12 |
| $^{64}$Zn | $^{65}$Zn, $^{63}$Zn, $^{62}$Zn, $^{64}$Cu, $^{61}$Cu, $^{60}$Cu, $^{57}$Ni, $^{58}$Co, $^{57}$Co, $^{56}$Co, $^{55}$Co, $^{56}$Mn, $^{52}$Mn, $^{48}$Cr, $^{44m}$Sc, $^{44}$Sc(i,m+g) | 17 |
| $^{63}$Cu | $^{61}$Cu, $^{58}$Co, $^{55}$Co, $^{52}$Mn | 4 |
| $^{59}$Co | $^{57}$Ni, $^{58}$Co(i,m+g), $^{58m}$Co, $^{57}$Co, $^{56}$Co, $^{55}$Co, $^{59}$Fe, $^{52}$Fe, $^{56}$Mn, $^{52}$Mn, $^{51}$Cr, $^{48}$Cr, $^{48}$V, $^{48}$Sc, $^{47}$Sc(i,c), $^{46}$Sc, $^{44m}$Sc, $^{44}$Sc(i,m+g), $^{43}$Sc, $^{47}$Ca, $^{43}$K, $^{24}$Na, $^{7}$Be | 222 |
| $^{27}$Al | $^{27}$Mg, $^{24}$Na, $^{22}$Na, $^{7}$Be | 186 |
| $^{19}$F | $^{18}$F | 1 |
| $^{12}$C | $^{11}$C | 1 |
| | Total | 979 |

The secondary radiation-induced absorbed doses were measured at 10 points on the W disc surfaces using the ID-10 dosimeter system, whose sensing detectors are made of radio-photoluminescent glass. The metering unit of the facility was calibrated in advance for the eventual results to correspond to the absorbed dose levels in biological tissue.

III. COMPUTATIONAL SIMULATION OF THE MEASURED REACTION RATES

The reaction rates were simulated by the LAHET Code System [5] using the LAHET and HMCNP codes. The high-energy (>20 MeV) hadron-nucleus interactions were simulated by LAHET code. The proton beam parameters were prescribed to conform to the results of determining the proton beam shape. The hadron-nucleus interactions were simulated in terms of the ISABEL model. The calculations were made allowing for multiple scattering of primary protons and for elastic scattering of >20 MeV neutrons. The HMCNP code was used to trace the slow ($E_n$<20 MeV) neutrons. The code system has generated the neutron and proton spectra in the experimental sample locations (see Fig. 3).

The simulated reaction rates may then be determined via integral multiplication of the calculated spectra by the respective reaction cross sections:

$$R_x = R_{n,x} + R_{p,x} = \sum_{i=n,p} \int \sigma_{i,x}(E) \cdot j_i(E) dE \quad (1)$$

Normally, the ENDF/B-type cross section libraries used extensively in the reactor physics include the neutron data estimates up to 20 MeV. Compilation of a cross section library above 20 MeV is, therefore, an urgent scientific task bearing on the revival of the ADS concepts as mentioned in the Introduction above. A few libraries of cross sections above 20 MeV (MENDL2, LA150, ENDF/B-VI(HE), WIND, and others) were compiled during the last decade. However, the highest energy of the libraries is normally 100-200 MeV and, besides, they need being experimentally verified in detail. Therefore, the necessary cross sections are expedient to accumulate not only using those libraries, but also with the help of the LAHET-type codes to generate the cross sections above 100 MeV and using the experimental reaction cross sections in the given energy range, if any.

That is why the present work accumulated the required proton and neutron excitation functions $\sigma_{i,x}(E)$ using:

- the MENDL2 [6] and MENDL2p [7] databases, which give the neutron and proton reaction excitation functions up to 100 MeV and 200 MeV, respectively;
- simulation of the excitation functions by LAHET code from 100 MeV to 800 MeV;
- available experimental data on the production cross sections of the respective reaction products (the EXFOR database).

Basing on the above dataset for a particular reaction, we can select the excitation function that would be consistent with all the data and agree satisfactorily with the experimental reaction rates at all measured points.

It should be noted, however, the experimental data include the production rates of some products that cannot be simulated using the MENDL database and the LAHET code. This relates to the production rates of metastable-state products, or ground-state nuclides when they can also be in metastable states, as well as to the product nuclides, whose production chain includes the metastable state transition to a nuclide outside a given decay chain fragment. Besides, the comparison with experimental did not include the occurrences of any strong correlation between the production rate of parent and daughter, as well as between the cumulative and independent production rates of a single nuclide. The $^{12}C \to {}^{11}C$, $^{19}F \to {}^{18}F$ reaction rates were not simulated either because of the lack of the appropriate MENDL data. The (n,γ)-type reactions were also excluded from the simulation because, contrary to others, they are not the threshold reactions and proceed mainly in the resonance and thermal neutron energy ranges, thus necessitating additional simulation of neutron moderation in the target environment.

Table 2 lists the product nuclides, whose production rates correspond to said cases.

Table 2. List of reaction rates measured, but not simulated.

| Nuclide | Products of non-simulated reactions |
|---|---|
| $^{209}$Bi | $^{206}$Bi(c), $^{200}$Tl(c) |
| $^{197}$Au | $^{198}$Au |
| $^{181}$Ta | $^{182}$Ta, $^{178m}$Ta, $^{180m}$Hf, $^{170}$Hf |
| $^{169}$Tm | $^{160m}$Ho (i,c) |
| $^{115}$In | $^{113}$Sn, $^{116}$In, $^{115m}$In(i,c) $^{114m}$In, $^{113m}$In(i,c), $^{110}$In, $^{115}$Cd, $^{111}$Ag, $^{110m}$Ag, $^{106m}$Ag, $^{105}$Ag, $^{101}$Rh, $^{99}$Rh, $^{99}$Nb(i), $^{89}$Zr |
| $^{93}$Nb | $^{90}$Y, $^{87m}$Y, $^{86}$Y(c) |
| $^{65}$Cu | $^{44m}$Sc, $^{44}$Sc(i) |
| $^{64}$Zn | $^{65}$Zn, $^{44m}$Sc, $^{44}$Sc(i) |
| $^{63}$Cu | - |
| $^{59}$Co | $^{58}$Co(i), $^{58m}$Co, $^{47}$Sc(c), $^{44m}$Sc, $^{44}$Sc(i) |
| $^{27}$Al | - |
| $^{19}$F | $^{18}$F |
| $^{12}$C | $^{11}$C |

The following two groups of reactions, whose cross sections are determined by different algorithms, can tentatively be singled out.

1. The reactions supported by sufficient experimental proton cross section data throughout the necessary energy range. The MENDL and LAHET excitation functions are, then, normalized and joined to each other, so that they would optimally describe the set of experimental points. In this case, the same normalization parameters were also applied to the neutron excitation functions that are not supported by any experimental points. We mean here the above-mentioned proceeding identity of the proton and proton reactions at high energies. It should be borne in mind, however, that, the MENDL data are probably very reliable at the energies adjoining the reactor range (this was also indicated above). This pattern was used to process the Au and Co data (except for the $^{24}$Na and $^{7}$Be data, which have been processed using only the experimental points for the reasons mentioned above).

2. The reactions supported by meager, or even none of, experimental data. In this case, where the comparison between experimental and calculated reaction rates shows systematic deviations of calculations from experiment, LAHET and MENDL were used to normalize the excitation functions in such a manner that the deviations would be minimized.

Fig. 4 exemplifies the usage of the above algorithms by the $^{59}Co \to {}^{55}Co$ reaction.

The differences between the calculated and experimental reaction rates were estimated using the squared deviation factor $<F>$ (calculated as $<F> = 10^{\sqrt{A}}$, where $A = <[\log(\sigma_{calc,i}/\sigma_{exp,i})]^2>$ [8]. The results of simulating the reaction rates are shown in Figs. 5-11 together with the experimental data.

The doses of the lateral W disc surfaces were calculated separately for the proton and neutron components to be the specific energy deposit in the biological tissue cells, and were then summed up. Fig. 11 shows the calculation results of dose rates together with experimental data.

## IV. COMPARISON BETWEEN EXPERIMENT AND CALCULATIONS

Figs. 5-11 show that most of the experimental results can be simulated to within a satisfactory accuracy (almost a half of the calculated data differ from experiment by less than 30%). This concerns the samples outside the target and the Al and Co samples on the rulers inside the targets at different distances from the target axis and from the beam hit point.

At the same time, the figures demonstrate some evident differences between calculations and experiment, namely,

1. The calculated $^{209}Bi(p,3n)^{207}Po$ and $^{209}Bi(p,4n)^{206}Po$ reaction rates in the first and eighth W discs are much underestimated compared with experiment. Since $^{207}$Po and $^{206}$Po cannot be produced in neutron reactions on $^{209}$Bi, we think that the discrepancy has arisen most probably from the fact that the LAHET code simulates rather inadequately the secondary proton production in backward direction in a hadron-nucleus cascade. Fig. 3 demonstrates the extremely low value of the simulated pro-ton flux on the surface of the first W disc (point S1). The fact is demonstrative also that the experiment-calculation agreement is usually good in the cases where the neutron component is of significance, which can be illustrated by the data on the rate of $^{24}$Na production from Al (see Fig. 8) for actually all versions of Al sample positioning along the target coordinates.

2. From Figs. 5, 6 and 7 it is seen that point S1 is characterize by a significant underestimation of the data calculated for not only the proton-induced reactions proper, but also the high-threshold (~100 MeV and higher) mixed-type reactions. Table 3 is the list of the reactions. It can be demonstrated that the above-mentioned point S1 proton anomaly alone cannot account for the observed discrepancy. Whereas, in the case of the proton-induced reactions of Po production in Bi, the experimental point S1-to-point S2 reaction rate ratio ($R_{S1}/R_{S2}$) is from 0.1 to 0.2 (thus suggesting the same ratio of the proton flux densities at the two points), it becomes quite evident that any adequate allowance for the proton component at point S1 would not lift the disagreement with the experimental data on the reactions listed in Table 3 because their rate ratio $R_{S1}/R_{S2}$ is 0.3-1.0. The disagreement can be explained by Fig. 3 (the third panel), which shows the neutron and proton spectra at points S1 and S2. The comparison between the neutron spectra at the two points indicates that, up to ~50 MeV, the S2 neutron flux density is about twice the S1 neutron flux density, in a good agreement with the $R_{S1}/R_{S2}$ ratio for the moderate-threshold reactions ($^{27}$Al→$^{24}$Na from Al, for instance). As neutron energy increases, however, the difference between the S1 and S2 flux densities rises, with the S1 flux vanishing at ~300 MeV. If the S1 neutron flux was actually the same as shown in Fig 3, the reactions with thresholds above ~200-300 MeV would have never been observed at point S1 (For example, $^{115}$In→$^{87}$Y, a ~400 MeV threshold). The significant experimental rates of the high-threshold reactions indicate that the simulated S1 neutron spectrum above ~100 MeV is underestimated because, possibly, the backward yield of high-energy neutrons and their transport through significant W thicknesses are simulated inadequately. A more definite conclusion will be drawn from processing the data, and analyzing the results, of the experiment with measuring the reaction rates on the lateral surface of a 40-mm thick single W disc irradiated by protons of the same energy.

Table 3. Comparison between experimental and calculated rates of high-threshold reactions at point S1.

| Reactions | Threshold, MeV | Underestimation factor |
|---|---|---|
| $^{201}$Pb, $^{200}$Pb, $^{200}$Tl(i,c) from $^{209}$Bi; $^{173}$Ta, $^{173}$Hf, $^{167}$Tm from $^{181}$Ta. | 50 – 100 | 2 – 5 |
| $^{191}$Pt from $^{209}$Bi; $^{101}$Pd, $^{100}$Rh (i,c) from $^{115}$In. | 100 – 200 | 10 – 50 |
| $^{100}$Pd, $^{97}$Ru, $^{96}$Tc, $^{95}$Tc, $^{90}$Nb, $^{87}$Y from $^{115}$In. | 200 and higher | 100 and higher |

Comparison between the experimental and calculated values of absorbed doses is indicative of significant differences throughout the studied interval of distances from the beam hit point. Surprisingly, from Fig. 12 it follows that the calculated doses are actually due to the proton component. This is emphasized by the fact that, the experimental dose on the surface of the first W disc is much higher than the calculated dose, contrary to most of other sample positions. This fact agrees with the above mentioned "Polonium" anomaly, which is particularly notable, as indicated above, on the surface of the first W disc.

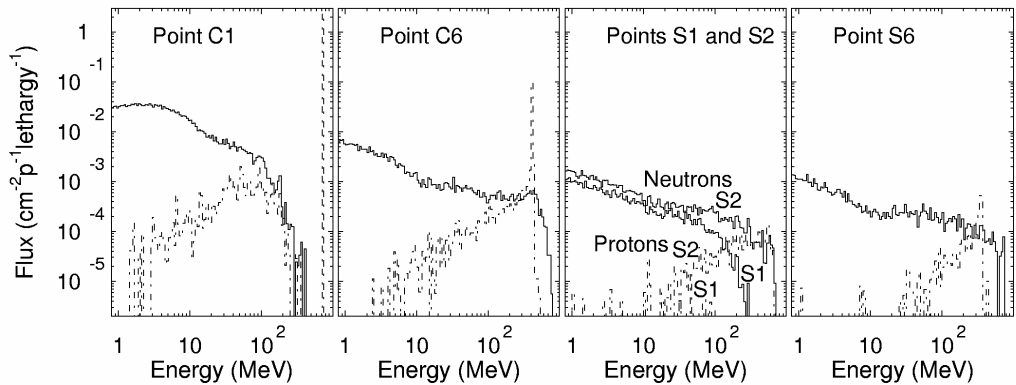

Fig. 3. The calculated neutron and proton fluxes at the selected points inside and outside the W-Na target. The solid and dashed lines show the neutron and proton fluxes, respectively. Point C1 means center of 1$^{st}$ W disc, C6 – center of 6$^{th}$ W disc, S1 – surface of 1$^{st}$ W disc, S2 – surface of 2$^{nd}$ W disc, S6 - surface of 6$^{th}$ W disc.

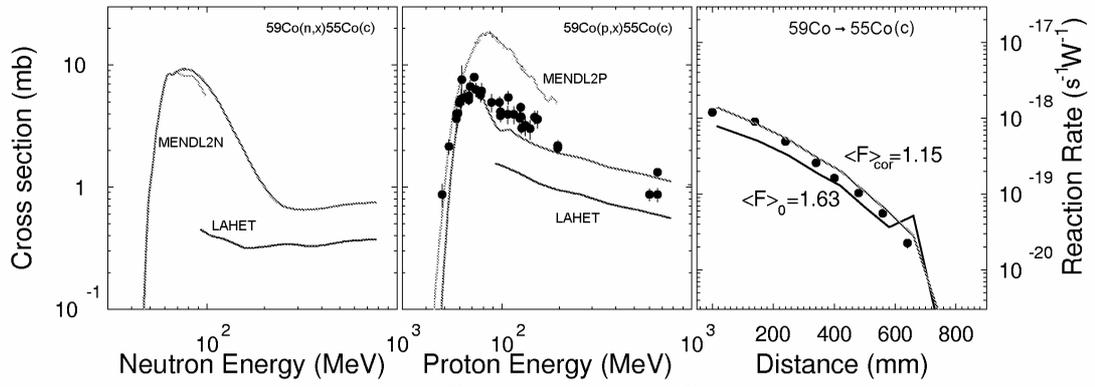

Fig. 4. The calculated cross sections and rates of $^{55}$Co production from $^{59}$Co.

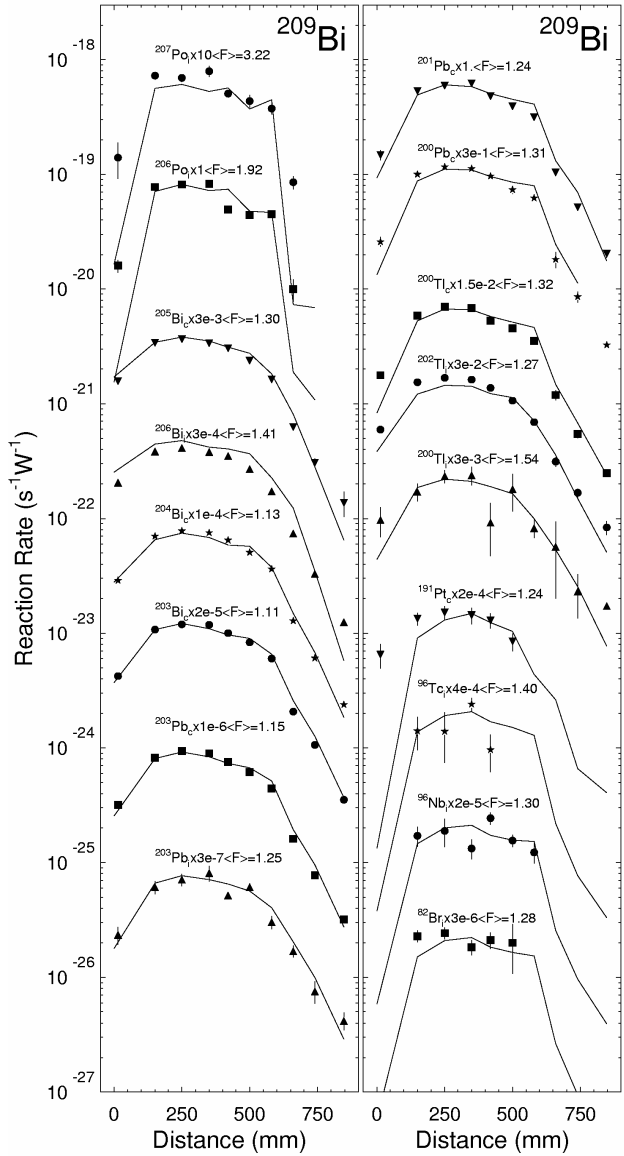

Fig. 5. The calculated and experimental reaction rates for $^{209}$Bi. The normalization factor and the mean squared deviation factor $<F>$ are presented for each reaction.

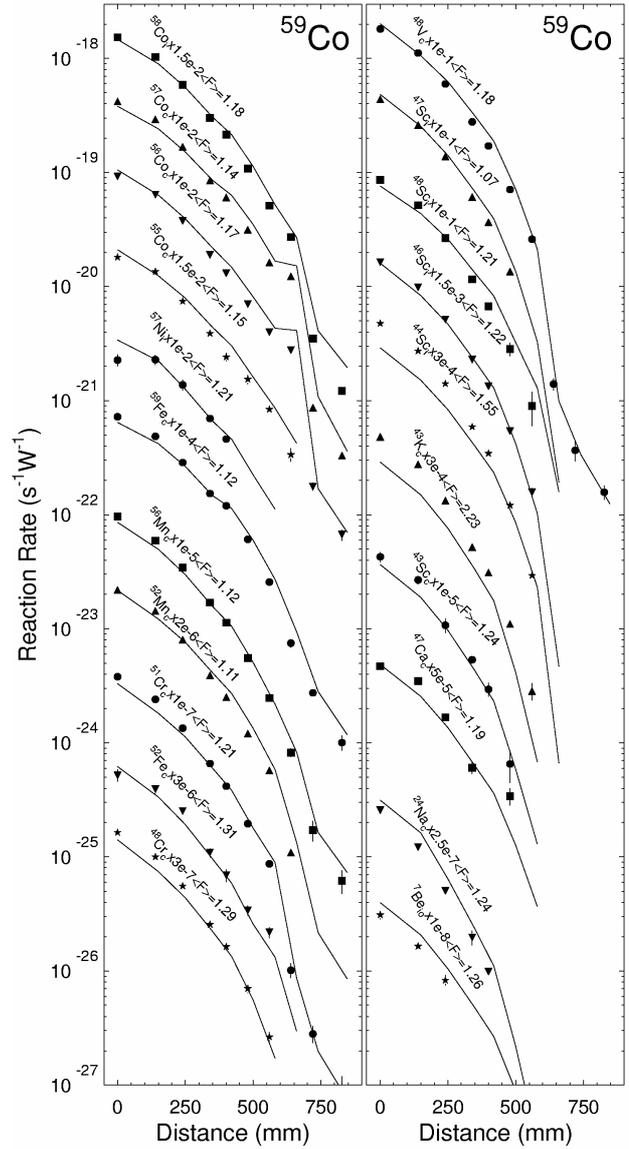

Fig. 6. The calculated and experimental reaction rates for $^{59}$Co. The normalization factor and the mean squared deviation factor $<F>$ are presented for each reaction.

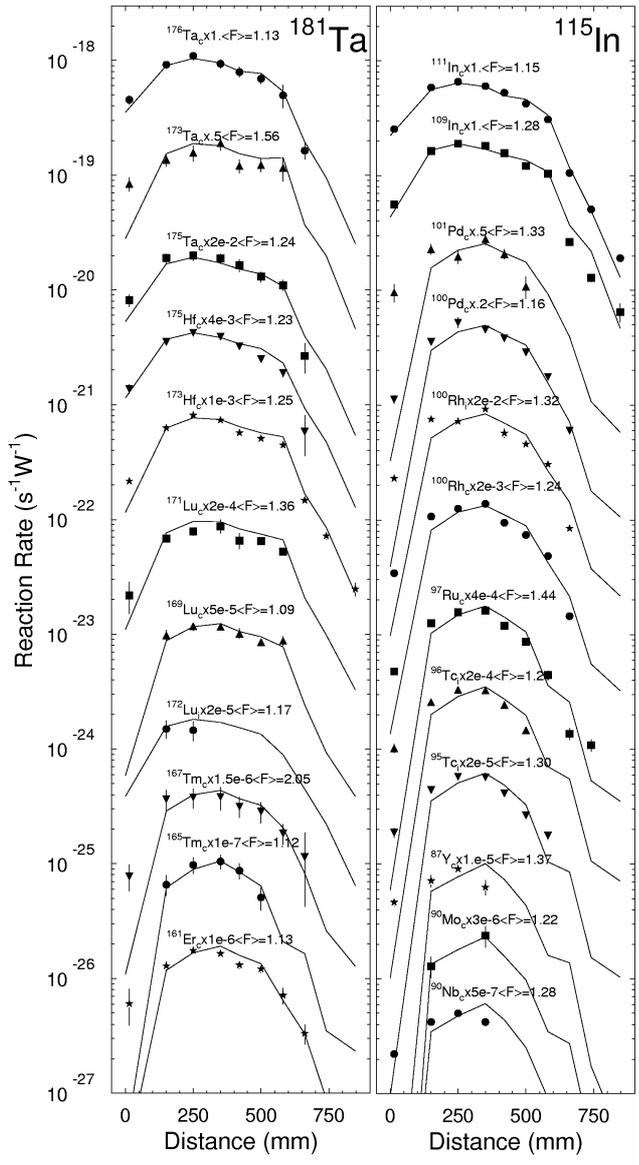

Fig. 7. The calculated and experimental reaction rates for $^{181}$Ta and $^{115}$In. The normalization factor and the mean squared deviation factor <F> are presented for each reaction.

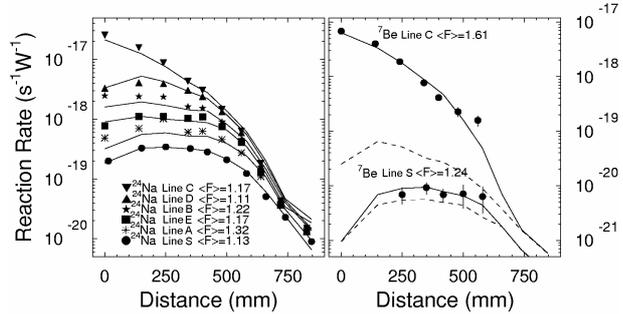

Fig. 8. The calculated and experimental reaction rates for $^{27}$Al. The dashed line is the contribution from the neutron component. The mean squared deviation factor <F> are presented for each reaction.

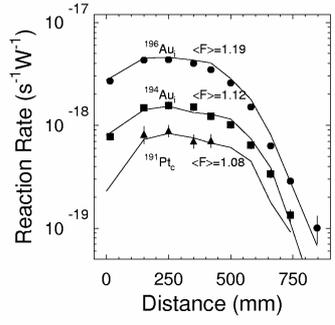

Fig. 9. The calculated and measured reaction rates for $^{197}$Au. The mean squared deviation factors <F> are presented for each reaction.

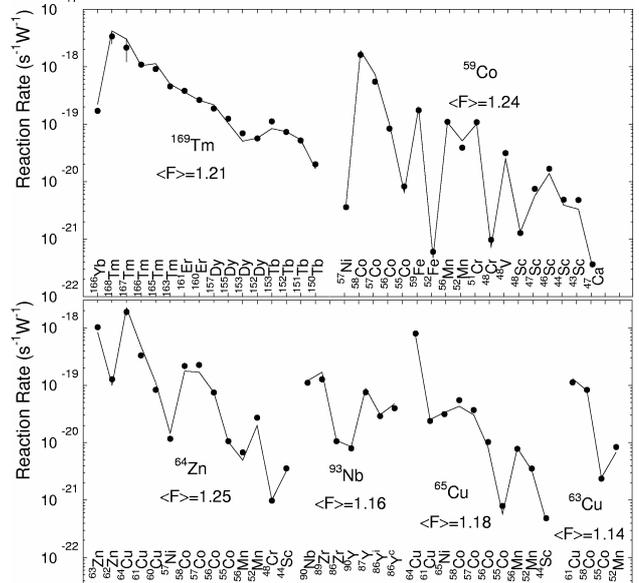

Fig. 10. The calculated and experimental reaction rates for $^{169}$Tm, $^{59}$Co, $^{64}$Zn, $^{65}$Cu, $^{63}$Cu, and $^{93}$Nb on the surface of the second W disc.

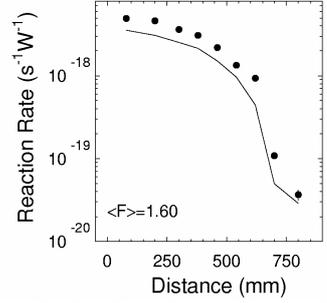

Fig. 11. The calculated and experimental rates of $^{22}$Na production in Na discs.

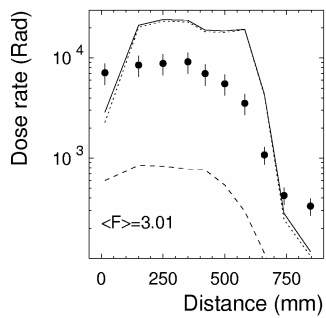

Fig. 12. The calculated and experimental absorption doses. The dashed line is the contribution from the neutron component.

It should be noted that the presented results of measuring and simulating the absorbed doses are but tentative because, until now, the respective researches were aimed mainly at the methodological aspect of the problem.

V. CONCLUSION

The results described above indicate that, with minor exclusions, the presented methods make it possible to obtain the excitation functions of high-energy threshold reactions that lead to a satisfactory description of the measured reaction rates. The functions can be used to unfold the neutron spectra at different points inside the ADS.

Further studies should nevertheless be aimed at perfecting the simulation of the production of secondary protons and high-energy neutrons, especially in the backward direction with respect to the beam. This is emphasized by the data on the rates of Po production from Bi and of high-threshold reactions sidewise of the beam entry to the target and by the tentative results of measuring the absorbed doses.

It should be noted that the satisfactory agreement in the reaction rates is somewhat conditional in the case of lacking reliable experimental data on excitation function for Bi, In, Ta, etc. This means that a significant uncertainty is possible to occur in the excitation functions. Therefore, some additional experiments must be designed to measure the production cross section of secondary product nuclei and the reaction rates in the integral experiments with other target types.

ACKNOWLEDGEMENT

The authors are indebted to Dr. F.E.Chukreev (Kurchatov Institute) for his assistance in working with the EXFOR database.

The work has been supported by International Science and Technology Center (ISTC) under Project#1145. In addition, the work was partially supported by the U.S. Department of Energy.